\documentclass[prb,preprint]{revtex4-1} 

\usepackage{amsmath}  
\usepackage{amsfonts} 
\usepackage{graphicx} 

\begin{document}

\title{Inexpensive Student-fabricated Solar Panels and 
Some Related Classroom Measurements}

\author{Nathan T. Moore}
\email{nmoore@winona.edu}
\author{Carl D. Ferkinhoff}
\email{cferkinhoff@winona.edu} 
\affiliation{Department of Physics, Winona State University, Winona, MN 55987}

\date{\today}

\begin{abstract}
We describe a procedure in which elementary education students fabricate 3-cell, $\approx5 ~\mathrm{Watt}$, $1.6~\mathrm{V}$ solar panels.  The panels are manufactured at a cost of about $\$6$ each.  The paper also describes a series of characterization activities that we have had our students complete with the panels once fabrication is complete.  Specifically, our students have determined:
the dependence of panel power on load resistance,
the variation of open-circuit panel voltage between different students' work,
and energy capture efficiency.
The activity also allows opportunities for discussion of manufacturing and process improvement which can provide unusual connections to engineering and technology learning outcomes.   

\end{abstract}

\maketitle 

\section{Introduction} 

Photovoltaic (Solar) panels are an increasingly common energy capture device that can be installed on residential and commercial rooftops.  Our students are generally excited about solar panels, and including them in classroom work can be very engaging.  However, a classroom set of panels can be quite expensive.  We've recently worked out a procedure in which students can fabricate panels in the classroom at a unit cost of about $\$6$ per $\approx5~\mathrm{Watt}$, $1.6~\mathrm{V}$ panel.

Once created, the panels can be used for a variety of authentic investigations, a few of which we describe at the end of the paper.  

The solar cells we have used \cite{ML_solar} are $3~\mathrm{inch}$ by $6~\mathrm{inch}$ rectangles of photovoltaic (PV) silicon.  When light shines on a cell, a potential difference of about $0.55~\mathrm{volts}$ appears between the front and back of the cell.  This voltage is relatively constant under different lighting conditions.  The electrical current the cell drives depends strongly on the light incident on the cell.  In some sense, a PV cell can be thought of as a light-controlled  current source with a constant voltage of $\approx0.55~\mathrm{volts}$.  

At earth's surface the intensity of sunlight energy on a clear day is about $1~\mathrm{kW/m^2}$.  If the cell is oriented to capture maximum solar flux, it presents an area of $18~\mathrm{in}^2\approx0.0116~\mathrm{m}^2$.  Accordingly, the maximum power available to a cell is $\approx11.6~\mathrm{Watts}$.  The vendor does not supply the specific chemistry of the cells, but states that cells produce $1.8~\mathrm{W}$ under standard solar conditions.  This efficiency, $1.8~\mathrm{W}/11.6~\mathrm{W}\approx15\%$, suggests that the cells are some sort of Crystalline Silicon photovoltaic material. \cite{NREL_PV_Cells}

\begin{figure}[h]
\centering
\includegraphics[width=\columnwidth]{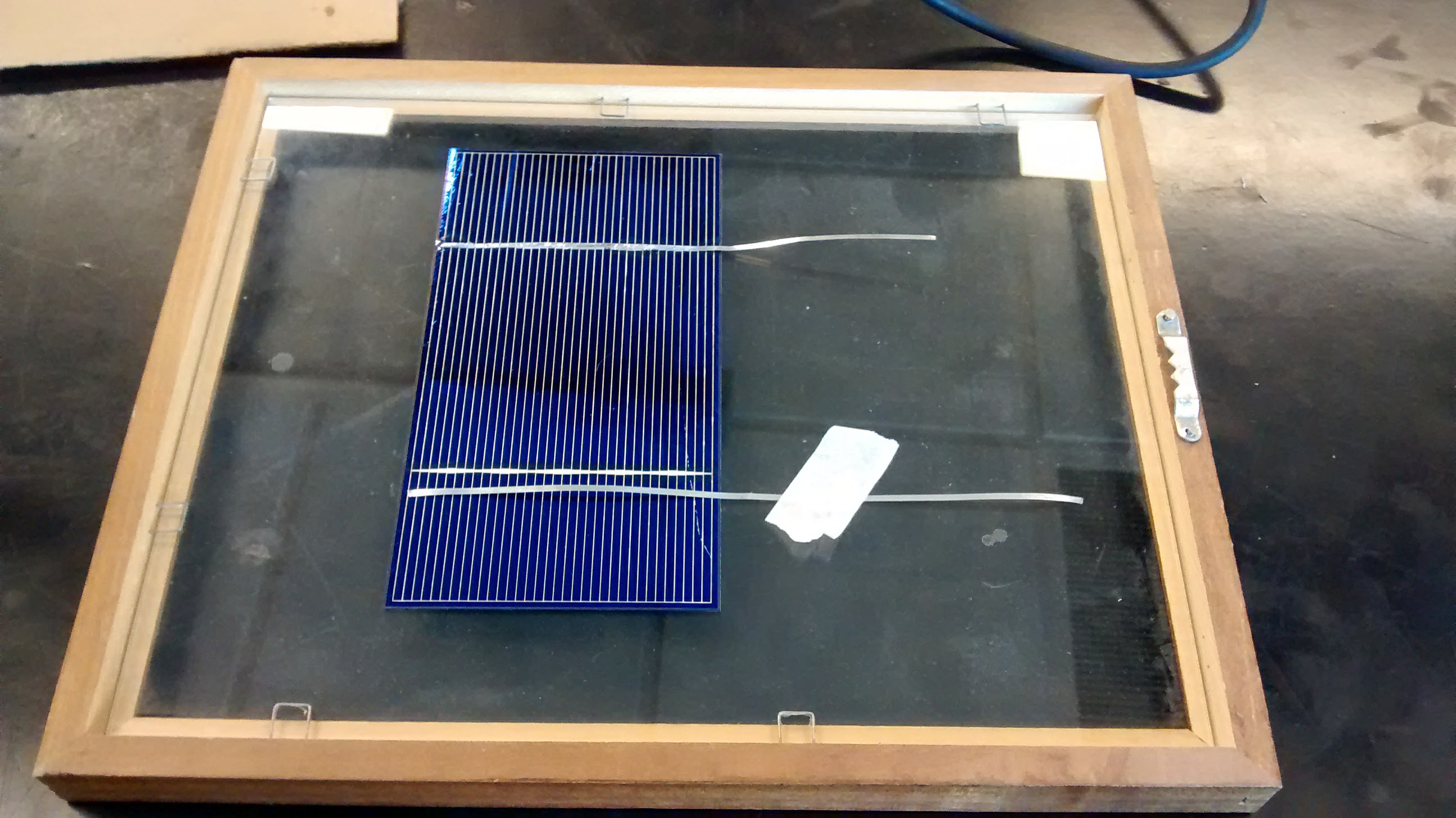}
\caption{
This photo shows the photovoltaic cells we used in solar panel fabrication.  Given the two long conductive traces on each cell, two pieces of tabbing wire are needed for each front-to-back series connection.  The top tabbing wire has been soldered in place.
}
\label{pv_cell}
\end{figure}

\section{Panel Fabrication and Costs}

PV cells are  fragile and the voltage a single cell produces is too low to be of practical value.  Commercial panels working at $12~\mathrm{volts}$
typically consist of $36$ individual PV cells which are wired in series to produce an open-circuit voltage of $\approx20~\mathrm{volts}$.  
Cells in this series circuit are electrically connected by flat, pre-tinned, ``tabbing'' wire which carries current from the back to the front of successive cells in a chain. 
I.e. current flows from the (positively charged) back of cell $i$ to the (negatively charged) front of cell $i+1$.  
Close inspection of a commercial panel reveals these tabbing wire connections.

Soldered chains of PV cells are encapsulated in a medium which will allow them to survive $25+$ years of exposure to the weather.  Generally, a piece of Iron-free tempered glass serves as the mounting surface for the ``sunny" side of the cells and an optically clear epoxy \cite{DOW_sylgard} serves as a bedding or potting medium that, applied to the shady side of the cells, holds them securely in place.  Dow Chemical maintains that their ``Sylgard" epoxy will retain $90\%$ light transmission over the $30$ year lifespan of a panel.  Iron-free, ``crystal-clear," glass gains the user another few percent of light transmission over standard tempered glass.  

Before pouring the epoxy encapsulant, terminal wires are soldered on to the overall top (negative) and bottom (positive) cells in the series chain.  
    
\subsection{Our Fabrication Procedure}

Written words are a poor substitute for physically soldering up a panel with a friend, and so we encourage you to watch a few videos on YouTube of people soldering up solar panels. \cite{1}

\begin{figure}[h]
\centering
\includegraphics[width=\columnwidth]{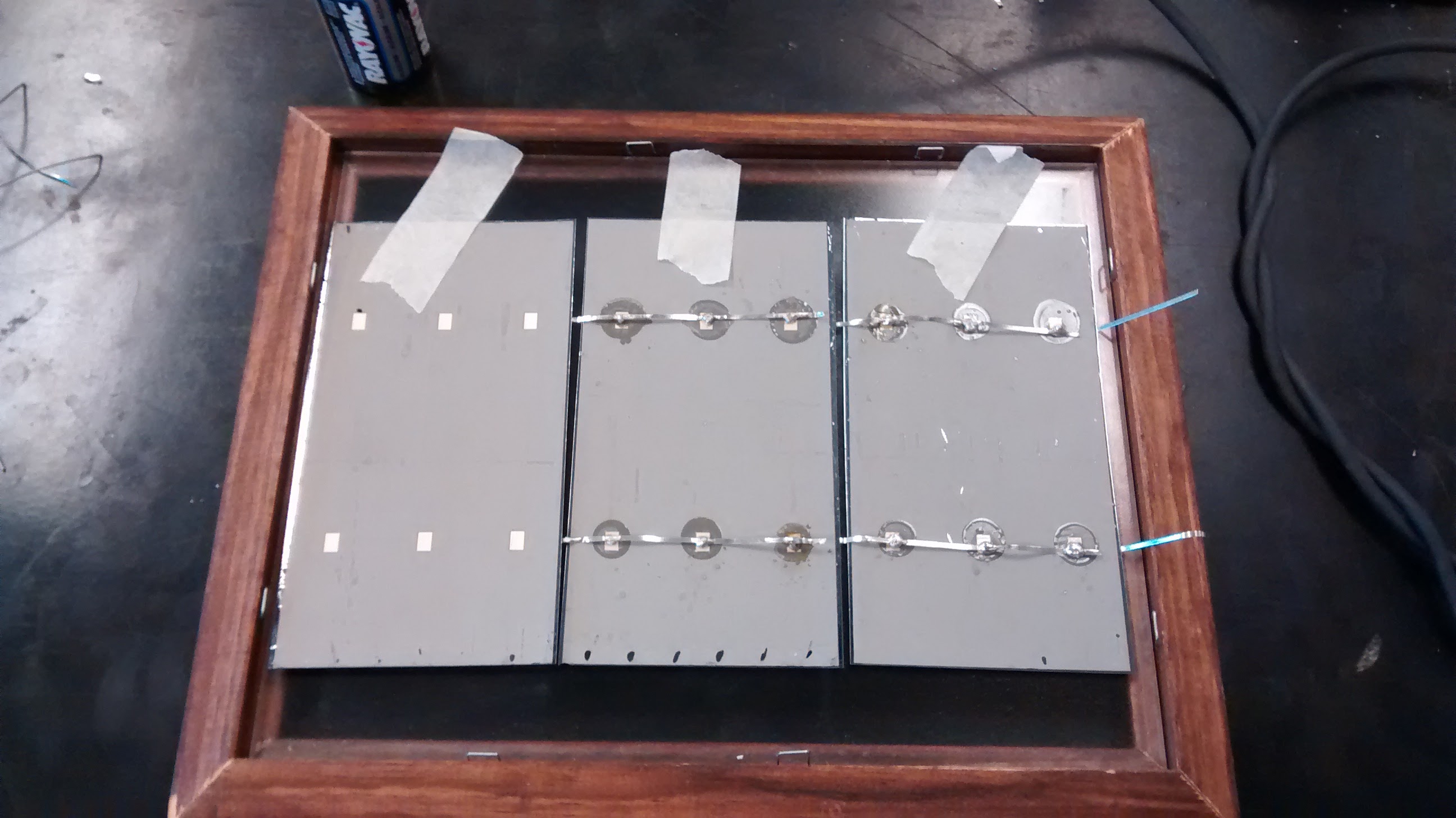}
\caption{
This photo shows a set of $3$ cells for which soldering is almost finished.  The loose tabbing on the right side of the picture will be the overall negative terminal of the panel and the $6$ open contact points on the left cell will be the overall positive contact.  Masking tape as an alignment tool was a student generated process improvement.
}
\label{almost_done}
\end{figure}

The procedure we've used with students follows:
\begin{enumerate}
\item
The activity generally falls after the initial “Batteries and Bulbs” unit of the ``CASTLE" E\&M curriculum. \cite{2}  Students are familiar with voltage, current, charge, and conductivity.  So far, we have successfully completed this activity with about $6$ classes of $28$ elementary education students (per class).  The $4$ credit class is an integrated lecture-lab that covers various topics in Physics and Chemistry.  While it is not a methods class, the instructors endeavor to use best-practice as understood by the Physics Education Research community.
\item 
A few days before assembly begins, students are given the homework of sourcing an $8$ by $10$ (inch) picture frame from a second-hand store and also watching a video about solar panel fabrication. \cite{3}
\item 
On Day 1, students remove the backing from their picture frame, clean the glass, and determine how many of the 3 by 6 cells will fit inside the frame.
\item 
Student then cut out $2$ pieces of tabbing wire, about $6$ inches long, and solder them to the two white traces on the face (blue, sunny side) of the cell.
\item 
The long ends of the tabbing wire fit underneath the adjacent cell, so that the panels are electrically connected ``head to tail," or negative (face) to positive (back).
\item It seems like the work turns out better if the tabbing wire is first soldered to the front of all three cells and then the overhanging wires are soldered to the backs of the adjacent cells.  
\item 
After the three cells are soldered together, standard $\approx24~\mathrm{ga}$ \cite{4} wire is soldered to the overall negative and positive ends of the array of cells.  Figuring out how to route the tabbing and where to attach the external contacts is a relevant time for the students to apply the idea of a ``closed conducting path." \cite{5}
\item
With wiring complete, the students can set the array of cells in the picture frame and gently brush on a coat of epoxy to bed the cells in place.  Uncured epoxy often leaks through the picture frame, and, if you do the epoxy application (bedding) with parchment paper protecting the table, you won't have to begin the next class meeting by chiseling the panel off of the table.
\end{enumerate}

\begin{figure}[h]
\centering
\includegraphics[width=\columnwidth]{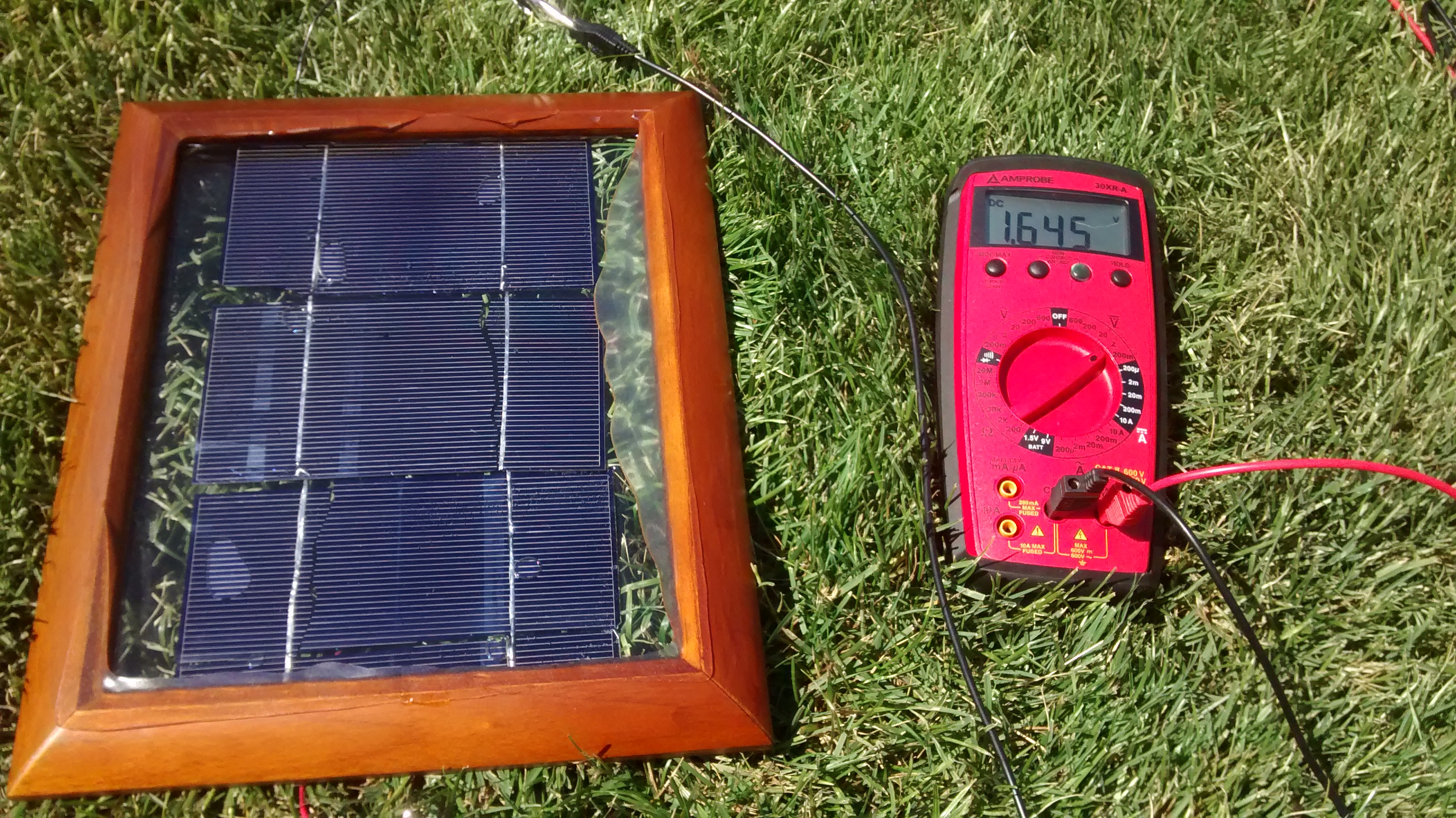}
\caption{
This photo shows a completed panel of $3$ PV cells.  The cells are held to the back of the picture frame glass by a coating of epoxy (which seeped through on the right side of the frame).  Note that while the cells cracked during assembly, the panel still produces about 
$0.55~\textrm{Volts}$ per cell.     
}
\label{pv_cell}
\end{figure}

In our experience, the first time your students fabricate a panel it will take about $2$ to $3$ hours of class/lab time, with an additional day to allow the epoxy to cure.  Our students have never soldered before this activity and we share about 15 soldering irons between 28 students.  With practice, fabrication becomes easier, and panel fabrication speeds up. After soldering on the first few pieces of tabbing wire, the remaining work can be  assigned as “homework” to be done in an open lab before the next class meeting. 

We have been surprised and pleased to learn how much our students love to solder!  In addition, as the fabrication process progresses, we have had very interesting student-driven conversations about the good and bad ways to solder up cell assemblies.

\subsection{Costs}
The materials you will need to source for this project average about $\$6.30/\mathrm{student}$. This unit cost could certainly be reduced.  Note, since this is a lab exercise, we are using generic ``picture-frame" glass and 2-part epoxy from the local home improvement store.  Our panels will likely be quite yellow after $30$ years of sun and they also probably won't survive a hailstorm.  For the sake of learning, we don't see either of these as problems.

Assuming $2$ classes of $30$ students, the materials you will need to source are estimated in Table \ref{bosons}.
\begin{table}[h!]
\centering
\begin{ruledtabular}
\begin{tabular}{l c c c}
Item 	&	Number	& Cost	& Total \\
 		& needed 	& per		& cost \\
\hline
$120$ Solar Cell kit, $3$ by $6$ inch cells, & $2$ & $\$100$ & $\$200$ \\
including tabbing wire and flux pen \cite{6} &&&\\
\hline
$1$ gallon clear epoxy resin  & $1$ & $\$70$ & $\$70$\\
 \& hardener \cite{7} &&&\\
\hline
$8$ by $10$ inch picture frame & $60$ & $\$1.50$ & $\$90$ \\
with clear glass \cite{8} &&&\\
\hline
Parchment (baking) paper, nitrile   & & 	$\$20$& 	$\$20$\\
gloves, disposable paintbrushes, &&&\\
glass cleaner, etc. &&&\\
\end{tabular}
\end{ruledtabular}
\caption{
This table gives a cost estimate for $60$ students to each build $3$-cell solar panels.  Construction also requires a classroom set of soldering irons, multimeters, and (optional) fume extractors.   
}
\label{bosons}
\end{table}

	At an estimated total cost of $\$380$ for $60$ student panels, this is roughly $\$6.30/student$.  Each cell is rated to produce $1.8~\mathrm{W}$ of power, which means that under ideal conditions, each student panel would produce 5.4W of power, or, they would produce power at $\$1.17/Watt$.  For context, at the consumer level, industrially manufactured panels $200~\mathrm{W}$ or more cost less than $\$1/\mathrm{Watt}$. \cite{9}

\subsection{Safety Considerations}
 
The tabbing wire for panels often comes pre-coated with solder, but it isn't clear if that solder is lead-free.  While it seems unlikely that lead solder will vaporize during the soldering process, lead is a contact hazard, so students should certainly wash their hands after finishing their work.  Additionally, the flux used in soldering can be a respiratory irritant for some people, so adequate room ventilation, perhaps with the addition of a charcoal-filter fan, \cite{fume_extractor} is a good idea.  Finally, soldering irons run at hundreds of degrees Celsius and there is the small risk of burns and/or melted synthetic clothing if students are careless with the irons.

\clearpage
\newpage

\section{Classroom Activities}

Once the epoxy has cured, the panels can be used for classroom activities.

\subsection{Series and parallel panel connections}  
All PV panels produce a “direct current” (DC) voltage.  Many PV systems use panel voltages of 12, 24, or 48 volts which are fed into an inverter, a device which outputs “normal” 110VAC (alternating current) electrical power for use in a residential or commercial setting.  With a sufficient number of panels wired in series, students can produce the $8$ to $12~\mathrm{VDC}$ potential that one needs to run a computer cooling fan, or a $12~\mathrm{VDC}$ inverter. 

When run as an open-inquiry activity, this is a challenging project for students to solve.  It integrates the notion of closed conducting paths, Kirchhoff's voltage loop rule, and trouble-shooting loose wires with a multimeter.  When the students figure out how to get the fan running on their own it is an meaningful accomplishment! 

\begin{figure}[h!]
\centering
\includegraphics[width=\columnwidth]{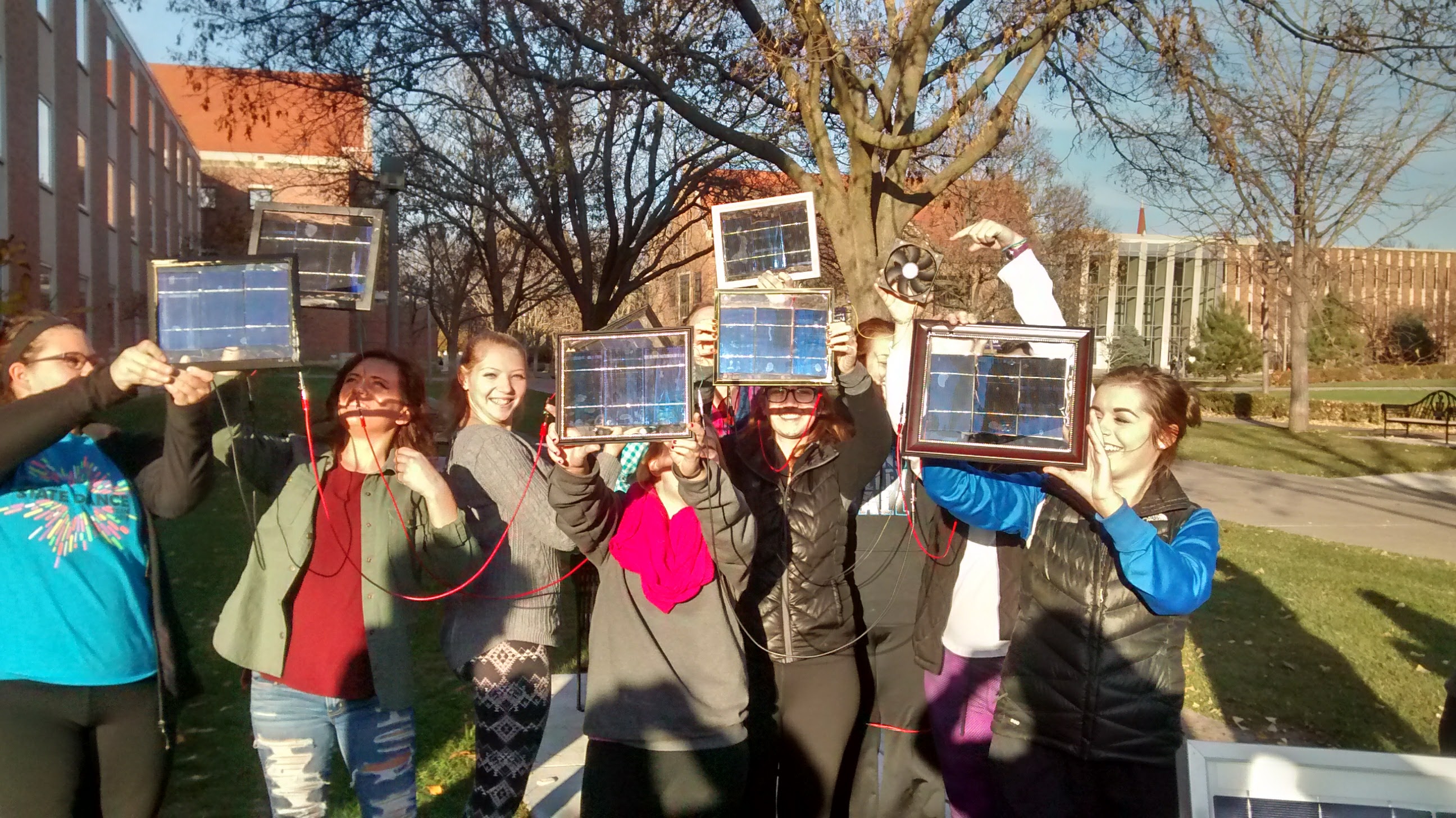}
\caption{
Future elementary school teachers powering a $12~\mathrm{V}$ CPU fan with mid November afternoon sunlight.  On the way outside one student earnestly stated, ``I've got to get this to work, I'm so invested!''
}
\label{running_fan}
\end{figure}

\clearpage
\newpage

\subsection{Manufacturing defects are interesting!}

Manufacturing is rarely discussed in science classes, but after a class creates $\approx30$ panels, substantial variation in panel output will be visible, and the students will likely have intuited significant knowledge about how to best fabricate a small PV panel. Statistics can be computed for panel output (voltage, current, and power), and meaningful reflection discussions can be had about the things we all learn while soldering.

\begin{figure}[h!]
\centering
\includegraphics[width=\columnwidth]{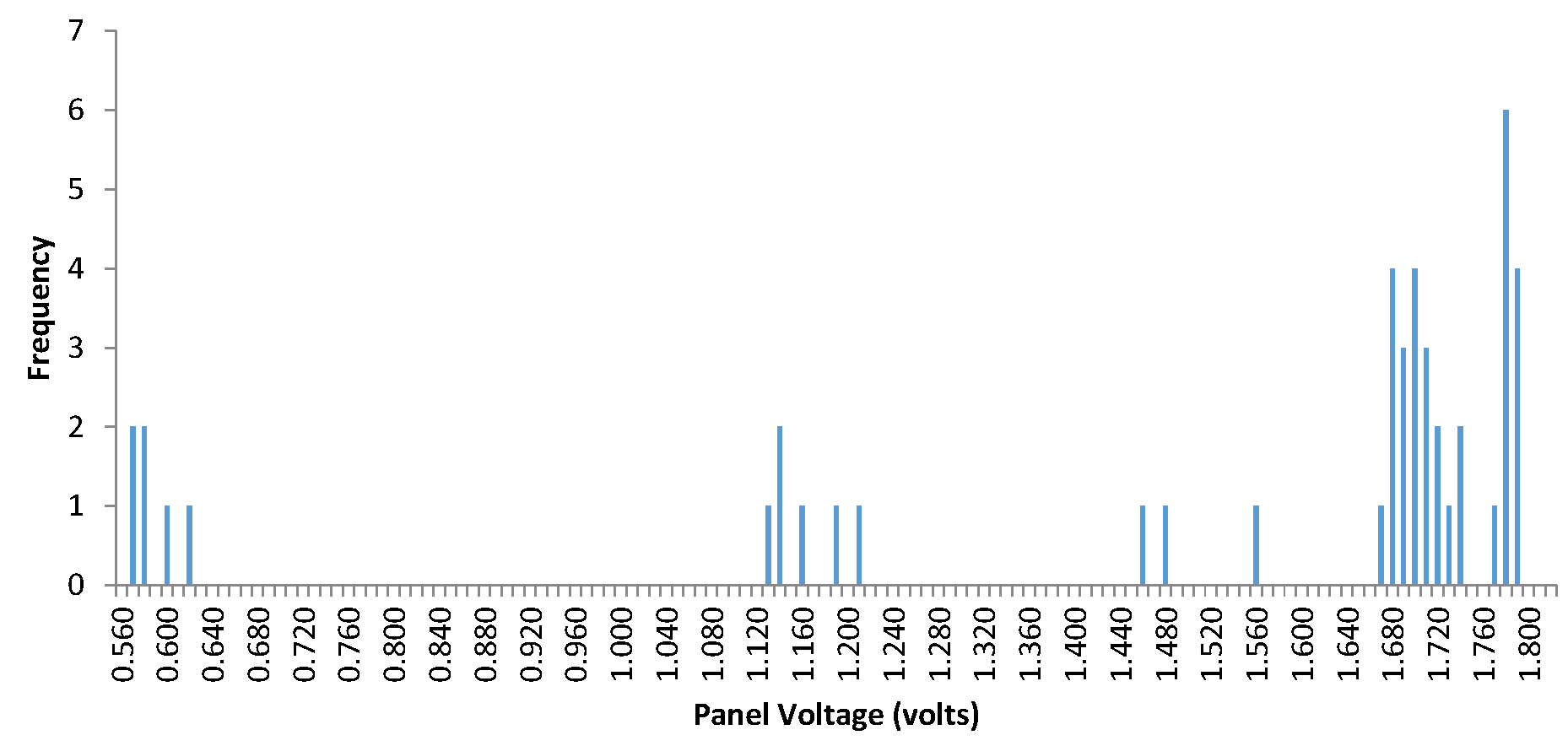}
\caption{This figure shows a histogram of open-circuit voltages, $\mathrm{V_{OC}}$, taken from $\mathrm{N}=46$ student-constructed solar panels under mixed outdoor sunlight.   Every data point binned into the histogram is a different panel. ``Open-circuit'' means the panels were disconnected from external load aside from a multimeter and effectively no current is flowing through the panel.  For the purpose of understanding manufacturing defects, note that while all the panels consisted of $3$ PV cells, there were clusters of apparent voltage at $\mathrm{V_{OC}}\approx0.6~\mathrm{V}$ and $\mathrm{V_{OC}}\approx1.2~\mathrm{V}$.  In almost all cases this occurred because students accidentally bridged the tabbing wire between the top and bottom of successive cells while soldering.  This accidental short circuit effectively removes the PV cell from the panel.
}
\label{panel_Voc_histogram}
\end{figure}

\begin{figure}[h!]
\centering
\includegraphics[width=\columnwidth]{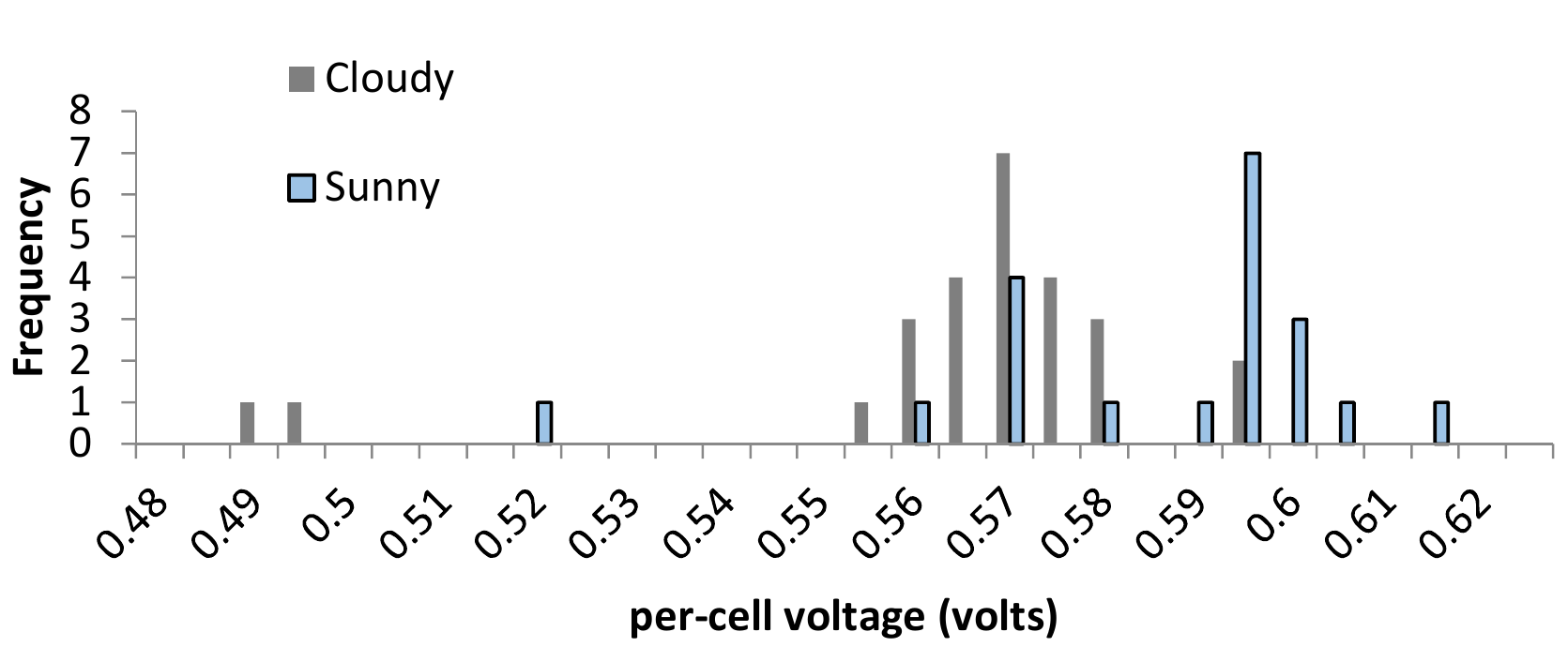}
\caption{
This figure uses the same raw data as figure \ref{panel_Voc_histogram}, however, the measurements are separated by classes that took data when it was sunny (blue, $\mathrm{N}=20$) or cloudy (grey, $\mathrm{N}=26$).  To make the histogram, the panel voltage, $\mathrm{V_{OC}}$, was divided by the number of functioning PV cells in the panel, eg, for a panel with $2$ functioning cells and a $\mathrm{V_{OC}}=1.14~\mathrm{V}$, the voltage used for the histogram would be $\frac{1.14~\mathrm{V}}{2}=0.57~\mathrm{V}$. While there is a subtle decrease in panel voltage when cloudcover increases, it is not enough to explain the voltage drop caused by a shorted out PV cell. 
}
\label{cell_Voc_histogram}
\end{figure}

Figures \ref{panel_Voc_histogram} and \ref{cell_Voc_histogram} show one such avenue for discussion.  After construction is finished, one of the first measurements we make is that of open-circuit voltage, $\mathrm{V_{OC}}$.  This is the panel voltage with only the multimeter as a load.  The impedance of a meter is at least $\mathrm{M}\Omega$ in size and so the panel voltage is measured under the condition of no current flow.  A histogram of student measurements, figure \ref{panel_Voc_histogram}, shows a primary voltage peak at about $\mathrm{V_{OC}}\approx1.7~\mathrm{V}$, which is what you might expect from a $3$ cell panel of $0.55~\mathrm{V}$ per cell.  

\begin{figure}[h!]
\centering
\includegraphics[angle=90,width=\columnwidth]{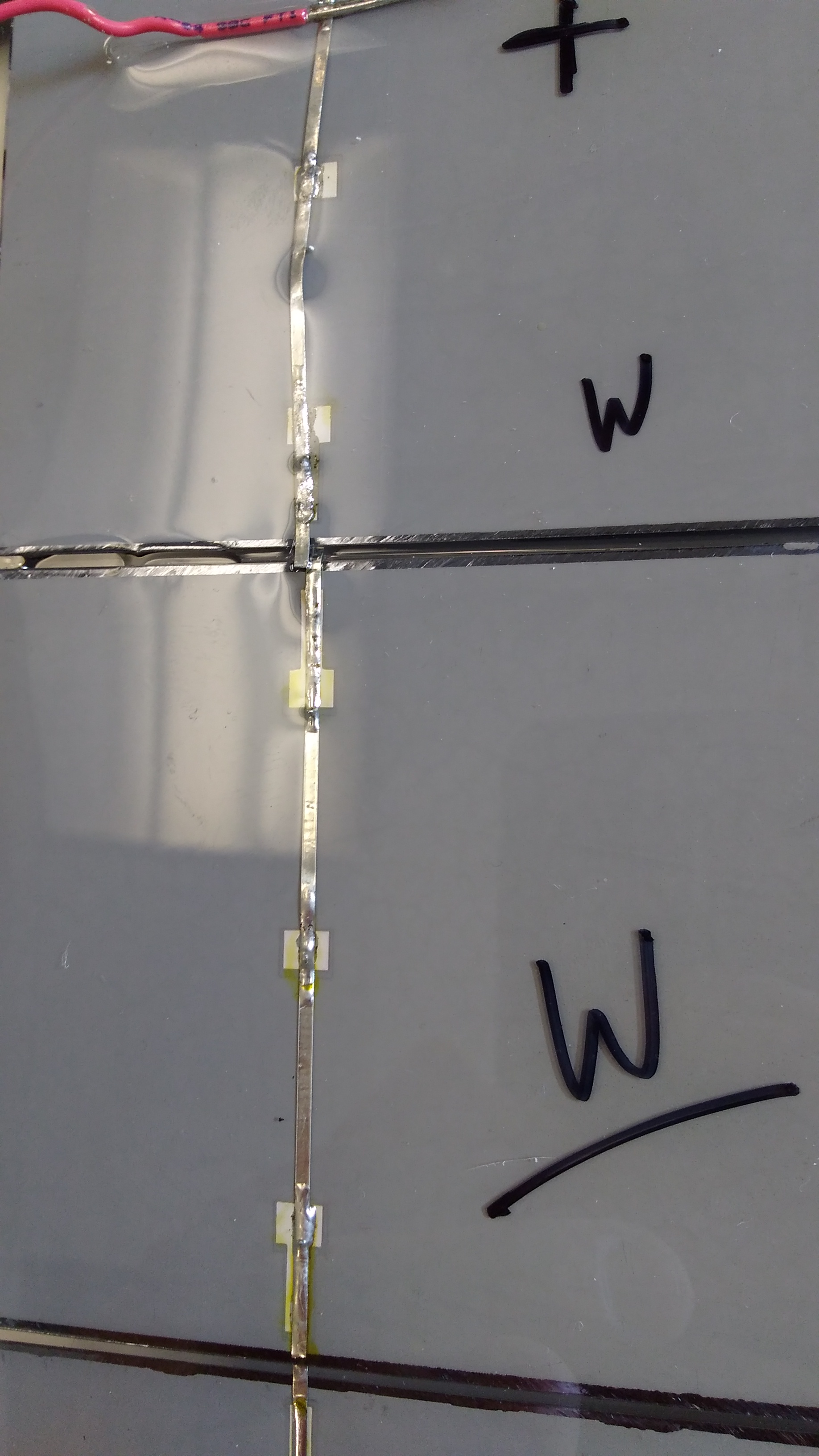}
\caption{This image shows a panel which produces an open-circuit voltage of 
$\mathrm{V_{OC}}\approx1.2~\mathrm{V}$.  On inspection, a student discovered that while the gap between cells on the right is ok, the tabbing wires on the left overlap, and the extra solder used effectively removes the leftmost cell from the panel's output voltage.
}
\label{too_much_solder}
\end{figure}

However, there are two smaller peaks, at $0.6~\mathrm{V}$ and $1.2~\mathrm{V}$.  After generating the histogram and wondering aloud with the class why not all of the panels have the same voltage, we suggest that students compare the soldering of expected ($1.7~\mathrm{V}$) and outlier ($0.6~\mathrm{V}$ and $1.2~\mathrm{V}$) panels.  Discussing student mistakes is always a delicate conversation, but we make a point of telling students how interesting and valuable mistakes are, as they allow us to learn.  In this situation, it is almost always the case that a student has bridged tabbing wire solder between the top and bottom of two adjacent cells, as figure \ref{too_much_solder} shows.    

To establish that it isn't just poor lighting that is making the multiple peaks in $\mathrm{V_{OC}}$, we have sometimes made figure \ref{cell_Voc_histogram} with the students, which shows per cell open-circuit voltage for the class set of panels.  The figure shows that while there is some variation in $\mathrm{V_{OC}}$ under different lighting conditions and panel orientations, it is not enough to account for the multiple peaks seen in figure \ref{panel_Voc_histogram}.

\begin{figure}[h!]
\centering
\includegraphics[width=\columnwidth]{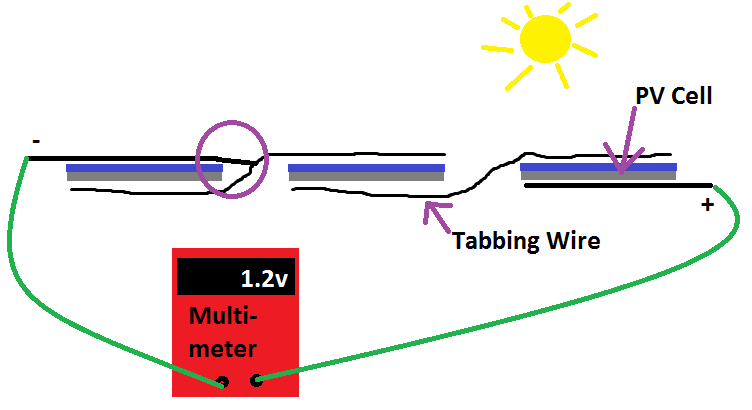}
\caption{
This cartoon shows the how overlapping tabbing wire can short out one of the cells in a PV panel. 
A Kirchhoff loop around the circuit shows that a charge must pass through two PV cells and the multimeter, but, because of extra solder at the circled location, the charge will bypass the third (left) PV cell and the multimeter will register $0.55~\mathrm{V}\cdot 2 \approx 1.1~\mathrm{V}$.
This is a common error that is easy to make while soldering and, when novel, is hard to decipher when students are testing panels. While frustrating, it provides a useful learning experience!
}
\label{too_much_solder_circuit}
\end{figure}

\clearpage
\newpage

\subsection{Maximum Power Point}
\begin{figure}[h]
\centering
\includegraphics[width=\columnwidth]{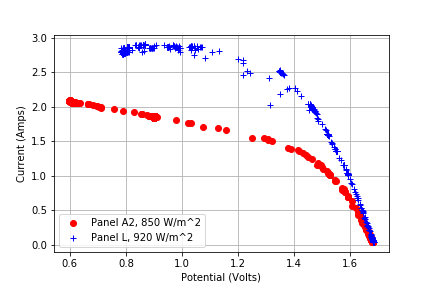}
\caption{
The voltage-current relationship for a $3$ cell panel which is oriented towards late-afternoon June Sun.  A Vernier Pyranometer \cite{pyranometer} gives the irradiance as $\approx920~\mathrm{W/m^2}$.  Orientation and sunlight were held constant and the load resistance was varied via a $25~\Omega$, $5~\mathrm{W}$ (linear variable) Potentiometer. \cite{pot}  Intercepts to the graph give an open-circuit voltage of $V_{OC}\approx 1.7~\mathrm{V}$ and short-circuit current of $I_{SC}\approx 2.8 ~\mathrm{A}$ for the undamaged panel, $L$.
}
\label{fig:iv}
\end{figure}

Solar panels exhibit an internal resistance (to current flow) that can be managed by “tuning” the load resistance that the panel drives. Qualitatively, the panel's voltage is largest when the leads are disconnected from a load, and the panel's current is greatest when the leads are short circuited by a wire.  This behavior is seen in figure \ref{fig:iv}, which shows some data from the lab-built panels already described.

The maximum power that can be produced by a panel is somewhere between these two limits and we use a potentiometer \cite{pot} as a variable load resistance to search for the panel's maximum power output.  Panel current and voltage are both functions of the load resistance, $R_L$, and by varying $R_L$, we can maximize the product, $P=IV$, and determine the maximum amount of power a panel can produce under given light conditions.  This maximum power point is also a function of (at least) the light incident on the panel, the temperature of the panel, the quality of tabbing wire connections, and physical damage done to the cells during assembly, and it seems, largely, a heuristic parameter.

\begin{figure}[h]
\centering
\includegraphics[width=\columnwidth]{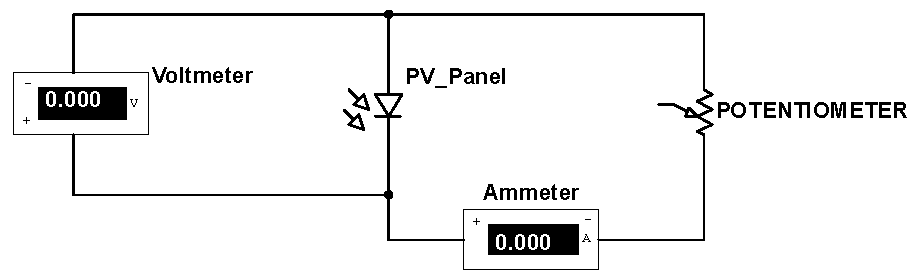}
\caption{
This circuit diagram shows the experimental setup used to measure panel current, voltage, and power.  The Voltmeter and Ammeter shown in the diagram are Vernier's High Current Sensor \cite{HCS-BTA} and Differential Voltage Probe. \cite{DVP-BTA}  Both sensors and Irradiation readings with a Pyranometer \cite{pyranometer} were collected with a LabPro or LabQuest Mini interface and Vernier's Logger Pro Software, running on a Windows PC.  Sensor readings were taken for 1-2 minutes at a sampling rate of about $10~\mathrm{Hz}$.
}
\label{fig:circuit_diagram}
\end{figure}

\begin{figure}[h]
\centering
\includegraphics[width=\columnwidth]{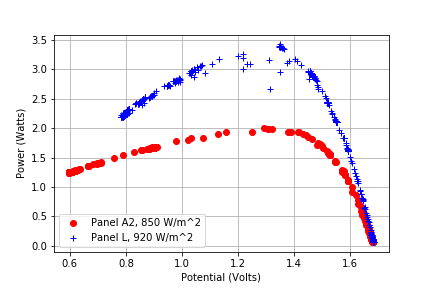}
\caption{
While the power output for the panel with damaged PV cells, A2, is about $2/3$ that of that produced by the clean panel, L, our students are continually surprised that the damaged panel is able to produce the same $V_{OC}$ as the clean panel.  The idea that voltage and energy are equivalent quantities might be the root of this misconception.  We've found that the question, ``Will the damaged panel produce the same voltage as the clean panel?" has been a worthwhile discussion topic.
}
\label{fig:pv}
\end{figure}

\begin{figure}[h]
\centering
\includegraphics[width=\columnwidth]{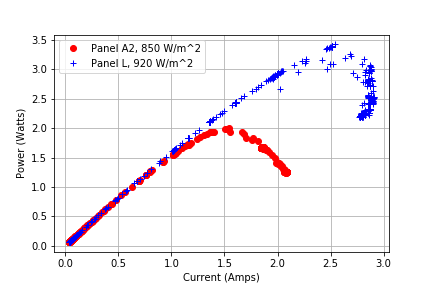}
\caption{It is interesting to see that for low current values, \textit{i.e.} large $R_L$ values, the current-power curves for the damaged, A2, and clean, L, panels are quite similar.  Logger Pro gave us a live readout of this figure and we rotated the potentiometer back and forth to search for the maximum power point.  Given the incomplete and somewhat random nature of our search (see particularly in the scatter of panel L values in this figure), our quoted maximums are probably uncertain at the $5\%$ level.     
}
\label{fig:pi}
\end{figure}

Variation of load resistance in residential and utility scale electric power generation is common, and in the photovoltaic industry the technique is sometimes described as 
``Maximum Power Point Tracking." \cite{mppt_review,mppt_inverter}
Using a decade resistor box \cite{decadeR} or potentiometer as a load for a student's panel is a way to model the search for the maximum power a panel can produce under given light conditions.  The setup we use is shown in figure \ref{fig:circuit_diagram} and the results for two panels are shown in figures \ref{fig:iv}, \ref{fig:pv}, and \ref{fig:pi}.

The data shown in these figures was captured by one of the authors at about 5pm on June 30, 2017 in Southeastern Minnesota with two 3-cell panels.  The first, panel ``L", was one of the highest power-producing panels we've seen.  The second, panel ``A2", is shown in figure \ref{pv_cell}, and uses PV cells with substantial cracks.  Both panels have access to all three PV cells, as they each produce $V_{OC}\approx 1.7~\mathrm{V}$, as can be seen in figure \ref{fig:iv}.

The sun came out late in the afternoon and the Pyranometer's Irradiation reading from the Sun was about $920~\mathrm{W/m^2}$ for panel L, and several minutes later, $850~\mathrm{W/m^2}$ for panel A2.  From our experience this  size variation in sunlight is normal over the course of an hour.  The difference in incident light, about $7.4\%$, does not account for the difference in power production -- panel A2's maximum power output was about $58\%$ of the power that panel L produced, and thus we see the difference that the internal resistance of a panel with cracked or damaged cells can make. 

The panels were oriented to present maximum area \cite{max_area} to incoming sunlight and then data was collected per the schematic in figure \ref{fig:circuit_diagram}. After initially scanning the general direction of the Sun to capture a maximum Solar Irradiation value we then used Logger Pro to record panel voltage and current values at $10~\mathrm{Hz}$ for about 1-2 minutes.  While the recording system was running, we rotated (wiggled) the potentiometer between high and low $R_L$ positions.  The data in figures \ref{fig:pv} and \ref{fig:pi} show that the maximum power panel L produced was  $3.43\mathrm{W}$ at a current of $2.54~\mathrm{A}$ and a voltage of $1.35~\mathrm{V}$.  Panel A2 had a maximum power of $1.99~\mathrm{W}$ at a current of $1.50~\mathrm{A}$ and a voltage of $1.30~\mathrm{V}$.  Given the incomplete and random way we rotated the potentiometers our maximum values are probably accurate to within $5\%$.
So, under these conditions the optimal load resistance is $0.53~\Omega$ for panel L, and $0.87~\Omega$ for panel A2, which, as figure \ref{pv_cell} shows, has substantial cracks and defects in the PV cells.  

Since power production depends strongly on $R_L$, it would be interesting to learn how much variation exists in the $R_L$ value that makes each panel produce maximum power.  We haven't yet had time to collect this data and think it would be an interesting topic to study further.

Each panel contains $3 \times \left( 3~\mathrm{inch} \times 6~\mathrm{inch} \right)$ PV cells, which present a total area of $54~\mathrm{inch^2}=0.0348~\mathrm{m^2}$ to capture solar energy.  If panel L was $100\%$ efficient, it would capture and generate a power of $0.0348~\mathrm{m^2}\cdot920~\mathrm{W/m^2}\approx 32~\mathrm{W}$.  Since the maximum power panel L is actually able to capture under this insolation is $3.43~\mathrm{W}$, it has a maximum efficiency of $3.43/32\approx10.5\%$.  Similarly, panel A2 exhibits an efficiency of $2~\mathrm{W}$ captured out of $29.6\mathrm{W}$ incident which is an efficiency of $\approx6.7\%$.   Most of the 3-cell panels with undamaged cells for which we've done this calculation have an efficiencies of about $9-11\%$.  

As mentioned, the vendor's specifications suggest that the cells are $15\%$ efficient.  A reminder, we used standard glass and generic epoxy in constructing our panels. The efficiency difference between the vendor's specification \cite{PVE_efficiency} and our results may be due to the economy materials we used to construct the panels.

\clearpage
\newpage

\subsubsection{Using a Halogen Lamp for Cloudy Days}

It is certainly possible to make the aforementioned characterization measurements indoors. Along the lines of others \cite{IOP_solar_article} we have sometimes (because of rain, early winter sunset, etc) used a Halogen work light as an artificial Sun equivalent.

\begin{figure}[h!]
\centering
\includegraphics[width=\columnwidth]{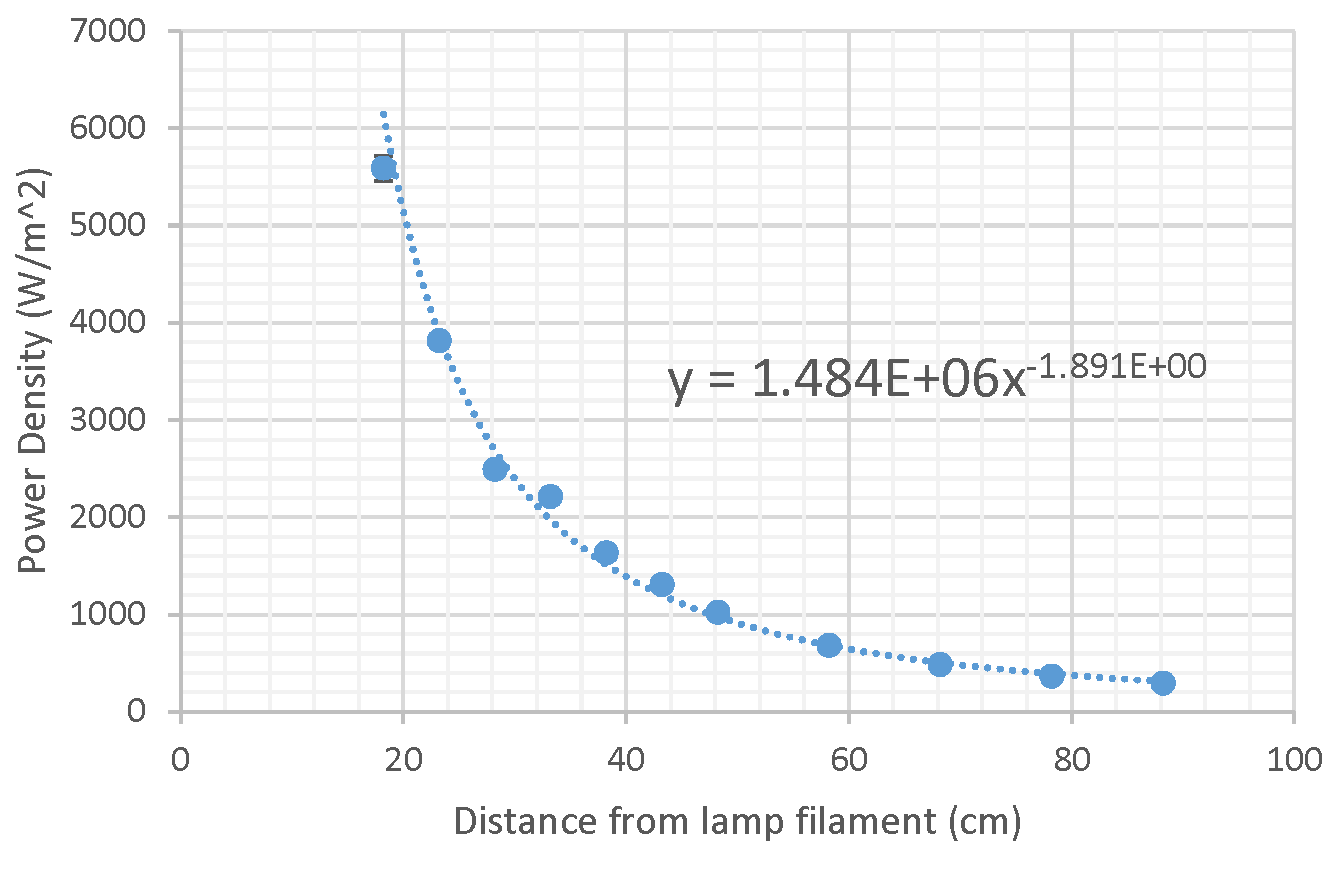}
\caption{Note, the horizontal scale in this graph is measured from the lamp filament, which is about $5.5~\mathrm{cm}$ from the glass shield which protects the filament.  The trendline crosses average solar insolation of $1000~\mathrm{W/m^2}$ at a distance of about $48~\mathrm{cm}$ from the bulb.  This position is equivalently a distance of about $42~\mathrm{cm}$ from the glass cover of the lamp.
}
\label{fig:Halogen}
\end{figure}

\begin{figure}[h!]
\centering
\includegraphics[width=\columnwidth]{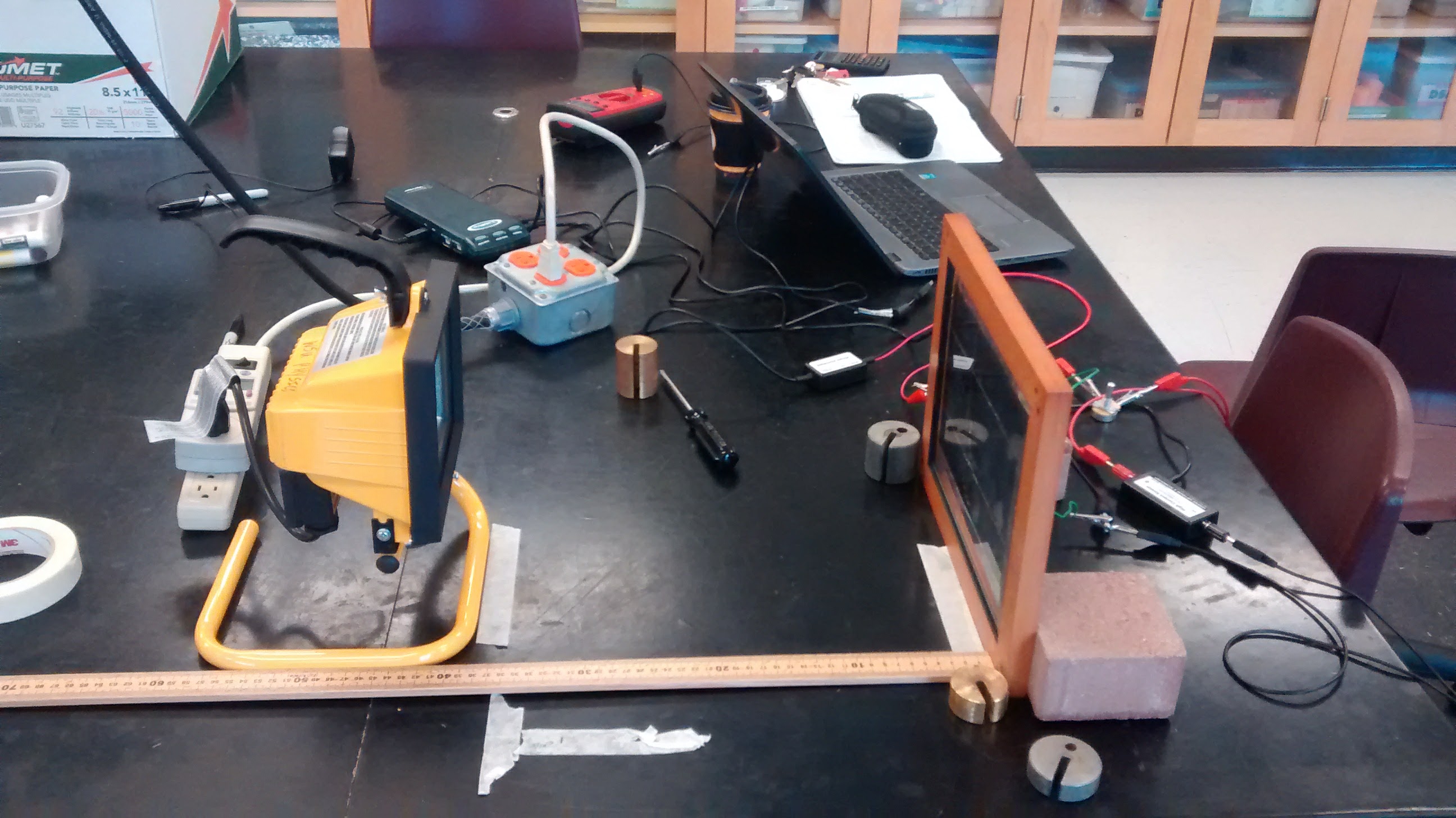}
\caption{
This photo shows a maximum power point setup for cloudy days.  Via a laser power meter we determined that a $500~\mathrm{W}$ Halogen work lamp placed about $42~\textrm{cm}$ from the PV panel provides a power of rougly $1~\mathrm{kW/m^2}$.  This method is only approximate though, as the hottest Halogen lamps only run at $\approx2700~\mathrm{K}$, much cooler (more yellow) than the spectrum of sunlight. 
}
\label{fake_sunlight}
\end{figure}

Figure \ref{fig:Halogen} shows power density (insolation) from a halogen work lamp as a function of distance from the lamp.  Via a Kill-A-Watt meter, the lamp was drawing $485\pm5~\mathrm{W}$ of power from a wall outlet.  Using a setup similar to that in figure \ref{fake_sunlight}, a Melles Griot 13PEM001 Laser Power Meter was used to measure the light power at different distances from the lamp.  The meter has a circular aperture of about $0.992\pm0.002~\mathrm{cm}$.  If we assume the full aperture is available to a power sensor, this corresponds to an area of about $0.773~\mathrm{cm^2}$.  We scaled the power registered by the meter by the aperture area to give the vertical scale shown in figure \ref{fig:Halogen}.  From the figure, if a panel is about 42cm from the glass (48cm from the filament), the incident light will be roughly equivalent to standard $1~\mathrm{kW/m^2}$ solar insolation.  

We should mention that the spectrum emitted by these Halogen lamps is quite yellow, $\approx2700~\mathrm{K}$.  We have not investigated the spectral absorption efficiency of the PV cells we're using. So, while the net power absorbed by a calorimeter at $48cm$ may be $1kW/m^2$, if the PV cells are (for example), less efficient at absorbing red photons, the panel efficiency will come in at a correspondingly lower number than expected.  A light source like the one described in \cite{IOP_solar_article} would allow one to investigate this interesting question further.

\clearpage
\newpage

\subsection{Light Transmission}
For the sake of minimizing cost, we have used home-improvement-store grade epoxy to bed the PV cells to the glass panel.  This, along with using cheap glass, is certainly not the commercial standard.  A transmission spectrum of the glass and cured resin would illustrate this reduction in the efficiency of solar energy capture.  We have not yet attempted this measurement with students but we think it would be a meaningful extension of the work.

\section{Conclusions}
While ``STEM" activities are, politically, quite popular right now, it is sometimes difficult to justify including ``Engineering" activities in a science content course.  It takes lots of time to build, test, revise and rebuild physical objects, and while there is great joy in physically creating new objects, the process doesn't always line up with the learning outcomes in science that our courses are meant to achieve.  Particularly because the discussion of ``defective" solar panels is so linked to Kirchhoff voltage loops, we feel this project is a meaningful addition to our course's $\approx2-3$ week discussion of voltage, current, power, and electric circuits.
We encourage you to consider teaching your students to solder!

\begin{acknowledgments}
This project was initially motivated by a Youtube video that was created and uploaded by Mr. Mark Patrick,  \url{<https://www.youtube.com/watch?v=_2UxOY_wpFo>}. We are grateful that he was willing to share his expertise on Youtube! We also thank Luke Zwiefelhoefer, Jennifer Zemke, Andrew Ferstl, Jon Mauser, and the Winona State students in Physics 180 who worked on this project with us.

\end{acknowledgments}


\begin{thebibliography}{99}
\bibitem{5} A “Closed Conducting Path” is an idea that's explicitly drawn out in the CASTLE E\&M materials.
\bibitem{8} Our local Salvation Army and Goodwill stores sell these for about $\$1.50$ per frame
\bibitem{decadeR}
Using a decade resistor box is actually a bad idea, as most contain only $1/4~\mathrm{W}$ or smaller resistors and the panels described have a theoretical maximum power of about $5.5~\mathrm{W}$.  Our panels are certainly sufficiently powerful to let the magic smoke out of a decade box.
\bibitem{max_area}
One easy way to align the panels such that they are capturing maximum solar flux is to tilt and rotate them so that the area of their shadow is maximized.
\end{thebibliography}
\end{document}